\documentclass{aa}

\usepackage{graphics}

\begin{document}

   \thesaurus{09     
              (08.06.2;  
               09.01.1;  
               09.03.1;  
               09.08.1;  
               09.09.1;  
               09.13.2;  
               13.19.3)} 
   \title{Study of the ammonia emission in the NGC\,6334 region}

   \author{A. Caproni\inst{1}, Z. Abraham\inst{1} and J. W. S. Vilas-Boas\inst{2}}

   \offprints{A. Caproni}

   \institute{Instituto Astron\^omico e Geof\'\i sico, Universidade de S\~ao Paulo,\\
              Av. M. St\'efano 4200, CEP 04301-904, S\~ao Paulo, SP, Brazil\
          \and
              Centro de Radio Astronomia e Aplica\c c\~oes Espaciais (CRAAE-INPE)\\
              CRAAM, Inst. Presbiteriano Mackenzie, R. Consola\c c\~ao 89, S\~ao Paulo, SP, Brazil. CEP 01302-000
             }

   \date{July 2000}

   \titlerunning{Ammonia in NGC\,6334 region}

   \authorrunning{A. Caproni et al.}

   \maketitle


   \begin{abstract}

   The region centered in the NGC\,6334\,I(N) radio continuum source was surveyed in an extension of 
   6\arcmin\ in right ascension and 12\arcmin\ in declination, in the NH$_3$(J,K) = (1,1) transition, 
   using the Itapetinga radio telescope. The spectra show non-LTE behavior, and gradients of velocity 
   and line-width were detected along the region. A detailed analysis of the spectra showed that the 
   surveyed region is composed of at least three overlapped sources related to regions that are in 
   different stages of star formation: NGC\,6334\,I, associated with an already known molecular bipolar 
   outflow, NGC\,6334\,I(N)w, the brightest ammonia source, coincidental with the continuum source 
   NGC\,6334\,I(N), and NGC\,6334\,I(N)e, weaker, more extended and probably less evolved than the 
   others. The physical parameters of the last two sources were calculated in non-LTE conditions, 
   assuming that their spectra are the superposition of the narrow line spectra produced by small 
   dense clumps. The H$_2$ density, NH$_3$ column density, kinetic temperature, diameter and mass 
   of the clumps were found to be very similar in the two regions, but the density of clumps is 
   lower in the probably less evolved source NGC\,6334\,I(N)e. Differences between the physical 
   parameters derived assuming LTE and non-LTE conditions are also discussed in this work. 

   \end{abstract}

%

\section{Introduction}

   The NGC 6334 region, located at a distance of 1.74 kpc from the Sun (Neckel \cite{nec78}), is a complex 
   of radio and infrared sources (Shaver \& Goss \cite{shgo70}; Haynes et al. \cite {hay78}; McBreen et al. 
   \cite{mcb79}). One of its components, NGC\,6334\,I(N), is also the strongest source in the NH$_3$(J,K) = 
   (1,1) transition (e.g., Schwartz et al. \cite{sch78}). It is also a strong continuum source at 400, 1000 
   and 1300 $\mu$m (Cheung et al. \cite{che78}; Gezari \cite{gez82}; Loughran et al. \cite{lou86}) but it 
   is below the detection limit at wavelengths smaller than 137 $\mu$m. It has an H$_2$O maser source, 
   CH$_3$OH masers (Moran \& Rodr\'\i guez \cite{moro80}; Kogan \& Slysh \cite{kosl98}; Walsh et al. 
   \cite{wal98}), six reddened sources at the JHK bands (Tapia et al. \cite{tap96}), a bipolar molecular 
   outflow in SiO and eight knots of H$_2$ emission (Megeath \& Tieftrunk \cite{meti99}). However, no 
   \ion{H}{ii} region associated to this source was detected.

   The source NGC6334 I, located approximately 2\arcmin\ south of NGC\,6334\,I(N), has H$_2$O, OH and 
   CH$_3$OH masers (Moran \& Rodr\'\i guez \cite{moro80}; Gaume \& Mutel \cite{gamu87}; Forster 
   \& Caswell \cite{foca89}; Menten \& Batrla \cite{meba89}), an ultra-compact \ion{H}{ii} region 
   (Rodr\'\i guez et al. \cite{rod82}; DePree et al. \cite{dep95}) and a cluster of stars detected at 
   the JHK bands (Tapia et al. \cite{tap96}). It was studied with high angular resolution in the 
   NH$_3$(J,K) = (1,1) transition by Jackson et al. (\cite{jac88}) (hereafter, JHH88), who interpreted 
   the observed structure as a circumstellar molecular disk rotating around a 30 $M_{\sun}$ O star. On 
   the other hand, CO and H$_{2}$ observations suggest that the structure arises from a molecular bipolar 
   outflow (Bachiller \& Cernicharo \cite{bace90}; Persi et al. \cite{per96}). Due to these characteristics, 
   it is probable that this region is in a more advanced stage of stellar formation than NGC\,6334\,I(N).

   Both regions have been studied in several transitions of $^{12}$CO, $^{13}$CO and CS (Kraemer \& 
   Jackson \cite{krja99}), the meta-stable transitions of the ammonia molecule, including NH$_3$(J,K) = 
   (1,1), (2,2), (3,3) and (6,6) [e.g., Schwartz et al. \cite{sch78}; Forster et al. \cite{for87}; 
   Vilas-Boas et al. \cite{vil88}; Jackson et al. \cite{jac88}; Kuiper et al. \cite{kui95}; Kraemer \& Jackson 
   \cite{krja95}, \cite{krja99}], and the non-meta-stable transition NH$_3$(J,K) = (2,1) (Kuiper et al. 
   \cite{kui95}). The physical parameters of these regions were always derived from the NH$_3$(J,K) = (1,1) 
   and (2,2) observations, assuming LTE conditions. In fact, the intensity anomalies in the nuclear quadrupole 
   hyperfine structure of the NH$_3$(J,K) = (1,1) spectrum are usually neglected, although they are observed 
   in many warm molecular clouds (Stutzki et al. \cite{stu82}; Batrla et al. \cite{bat83}; Stutzki et al. 
   \cite{stu84}). 

   Matsakis et al. (\cite{mat77}) interpreted the anomalous spectrum as the consequence of non-thermal 
   population in the hyperfine states, induced by selective trapping in the hyperfine transitions of 
   NH$_{3}$(J,K) = (2,1) $\rightarrow$ (1,1). This effect is relevant only when the width of the hyperfine 
   lines lie between 0.3 and 0.6 km s$^{-1}$. Due to this restriction, they assumed that the molecular cloud 
   is formed by clumps, each one producing a narrow line spectrum, so that the observed spectrum would be the 
   superposition of individual clump spectra. Based on this model, Stutzki \& Winnewisser (\cite{stwi85}) 
   (hereafter, SW85) elaborated a numerical algorithm to calculate the brightness temperature ratios between 
   the hyperfine satellites and the main line in the NH$_3$(J,K) = (1,1) spectrum, and also the brightness 
   temperature ratio between the (2,2) to the (1,1) main lines, as a function of the NH$_3$ column density, 
   H$_2$ density and kinetic temperature of each clump. 

   Gaume et al. (\cite{gau96}) questioned this model, based on observations of the NH$_3$(J,K) = (1,1) 
   absorption spectra towards the continuum source DR21. Their argument was based on an anti-correlation 
   in the degree of LTE departure of the inner and outer hyperfine components in different regions of the 
   cloud. They proposed instead, inflows and outflows of matter to explain the observations, but did not 
   present any quantitative calculation to support their claim.

   In this work, we used our high quality observations of the NGC 6334 region in the NH$_3$(J,K) = (1,1) 
   transition, which allowed the precise determination of the intensity anomalies in the hyperfine 
   structure, to obtain with good reliability its physical parameters assuming non-LTE conditions. 
   The data allowed us also to separate in the spectra the contribution of at least three different 
   sources: NGC\,6334\,I, NGC\,6334\,I(N)w and NGC\,6334\,I(N)e.

\section{Observations}

   The observations were made during July 1996 using the 13.7 m radome enclosed radio
   telescope of the Itapetinga Radio Observatory\footnote{Operated by CRAAE, Centro de 
   Radio Astronomia e Aplica\c c\~oes Espaciais, S\~ao Paulo, Brazil}. The receiver 
   front-end consisted of a circularly polarized corrugated horn connected to a cooled HEMPT. 
   An acusto-optical spectrometer was used in the back-end, with spectral resolution and
   total bandwidth of 70 kHz and 41 MHz respectively. The total system temperature was about 
   200 K. The angular resolution of the radio telescope at the frequency of NH$_3$(J,K) = (1,1) 
   transition was 4.2\arcmin. The observations were made using the ON-OFF total-power technique, 
   switching between positions every 20 s with amplitude of 20\arcmin\ in azimuth, enough to guarantee 
   that there was no source in the OFF position. The effective integration time of each observation 
   was 210 s. A 15 K noise source and a room temperature load were used in the calibration to obtain 
   the gain and to correct for atmospheric attenuation (Abraham \& Kokubun \cite{abko92}). The continuum 
   point source Virgo A was used as a primary calibrator. The NGC 6334 region was mapped in the 
   NH$_3$(J,K) = (1,1) transition, covering 6\arcmin\ in right ascension and 12\arcmin\ in declination, 
   with a spacing of 2\arcmin. The central position of the map, with equatorial coordinates $\alpha(1950) 
   = 17\rm^{h} 17\rm^{m} 32\rm^{s}$ and $\delta(1950) = -35\degr 42\arcmin$, was observed several times 
   during the mapping period and was used as secondary calibrator. This position was also observed in the 
   NH$_3$(J,K) = (2,2) transition. 

   A polynomial baseline was subtracted from the observations and the five Gaussian functions were fitted 
   to the hyperfine NH$_3$(J,K) = (1,1) spectra. The velocity separation between the Gaussians was fixed 
   and the line-width of the hyperfine satellites constrained to be equal to the main line. This last 
   assumption is well justified, since the optical depth of the main line is small ($\tau \sim 1.4$). 

   The results from the observations are presented in Table 1. The first column shows the position 
   in the map relative to the central coordinate, columns 2-6 exhibit the antenna temperature of 
   each transition corrected for atmospheric attenuation, columns 7 and 8 the velocity and line-width 
   of the main line and the last column the number of observations made in each position. The 
   NH$_3$(J,K) = (1,1) spectra of some points of special interest, as well as the NH$_3$(J,K) = (2,2) 
   spectrum of the central position are shown in Figure 1. 

   \begin{table*}
      \caption[]{Parameters obtained from the Gaussian fitting for each position in the map.}
      \scriptsize
      \begin{tabular}{ccccccccc}
         \hline \smallskip
         ($\Delta\alpha,\Delta\delta$)$^{\mathrm{a}}$ & T$\rm_A$(F=1$\rightarrow$0) & T$\rm_A$(F=1$\rightarrow$2) & T$\rm_A$($\Delta$F=0) & T$\rm_A$(F=2$\rightarrow$1) & T$\rm_A$(F=0$\rightarrow$1) & v$\rm_{LSR}$$^{\mathrm{b}}$ & $\Delta$v$\rm_{FWHM}$ & \#$^{\mathrm{c}}$ \\
         (arcmin) & (K) & (K) & (K) & (K) & (K) & (km s$^{-1}$) & (km s$^{-1}$)   \\
         \hline
\smallskip
(-2,-2) & 0.17~$\pm$~0.04 & 0.27~$\pm$~0.04 & 0.67~$\pm$~0.04 &	0.24~$\pm$~0.04 & 0.26~$\pm$~0.04 & -5.09~$\pm$~0.12 & 6.00~$\pm$~0.13 & 4 \\
\smallskip
(-2,0)  & 0.15~$\pm$~0.03 & 0.24~$\pm$~0.03 & 0.54~$\pm$~0.03 &	0.21~$\pm$~0.03 & 0.20~$\pm$~0.03 & -4.40~$\pm$~0.12 & 4.28~$\pm$~0.12 & 4 \\
\smallskip
(-2,2)	& 0.09~$\pm$~0.03 & 0.13~$\pm$~0.03 & 0.43~$\pm$~0.03 &	0.15~$\pm$~0.03 & 0.14~$\pm$~0.03 & -3.82~$\pm$~0.12 & 3.06~$\pm$~0.13 & 3 \\
\smallskip
(0,-4)	& 0.15~$\pm$~0.04 & 0.23~$\pm$~0.04 & 0.39~$\pm$~0.04 &	0.12~$\pm$~0.04 & 0.11~$\pm$~0.04 & -4.66~$\pm$~0.12 & 6.29~$\pm$~0.12 & 4 \\
\smallskip
(0,-2)	& 0.30~$\pm$~0.03 & 0.43~$\pm$~0.03 & 0.86~$\pm$~0.03 &	0.33~$\pm$~0.03 & 0.32~$\pm$~0.03 & -4.98~$\pm$~0.12 & 5.57~$\pm$~0.12 & 2 \\
\smallskip
(0,0)	& 0.41~$\pm$~0.01 & 0.62~$\pm$~0.01 & 1.41~$\pm$~0.01 &	0.53~$\pm$~0.01 & 0.52~$\pm$~0.01 & -4.66~$\pm$~0.12 & 4.26~$\pm$~0.12 & 19 \\
\smallskip
(0,2)	& 0.30~$\pm$~0.04 & 0.51~$\pm$~0.04 & 1.05~$\pm$~0.04 &	0.42~$\pm$~0.04 & 0.43~$\pm$~0.04 & -4.21~$\pm$~0.12 & 4.06~$\pm$~0.12 & 2 \\
\smallskip
(0,4)	& 0.15~$\pm$~0.03 & 0.24~$\pm$~0.03 & 0.62~$\pm$~0.03 &	0.17~$\pm$~0.03 & 0.29~$\pm$~0.03 & -3.36~$\pm$~0.12 & 3.11~$\pm$~0.12 & 4 \\
\smallskip
(2,-4)	& 0.10~$\pm$~0.04 & 0.22~$\pm$~0.04 & 0.44~$\pm$~0.04 &	0.21~$\pm$~0.04 & 0.21~$\pm$~0.04 & -5.98~$\pm$~0.12 & 5.37~$\pm$~0.12 & 2 \\ 
\smallskip
(2,-2)	& 0.27~$\pm$~0.04 & 0.39~$\pm$~0.04 & 0.86~$\pm$~0.04 &	0.34~$\pm$~0.04 & 0.34~$\pm$~0.04 & -5.29~$\pm$~0.12 & 4.57~$\pm$~0.12 & 2 \\
\smallskip
(2,0)	& 0.38~$\pm$~0.02 & 0.54~$\pm$~0.02 & 1.28~$\pm$~0.02 &	0.47~$\pm$~0.02 & 0.48~$\pm$~0.02 & -4.74~$\pm$~0.12 & 4.09~$\pm$~0.12 & 6 \\
\smallskip
(2,2)	& 0.34~$\pm$~0.06 & 0.32~$\pm$~0.06 & 1.06~$\pm$~0.06 &	0.47~$\pm$~0.06 & 0.43~$\pm$~0.06 & -4.43~$\pm$~0.12 & 4.21~$\pm$~0.13 & 1 \\
\smallskip
(2,4)	& 0.16~$\pm$~0.03 & 0.28~$\pm$~0.03 & 0.59~$\pm$~0.03 &	0.25~$\pm$~0.03 & 0.23~$\pm$~0.03 & -3.71~$\pm$~0.12 & 3.53~$\pm$~0.13 & 3 \\
\smallskip
(2,6)	& 0.13~$\pm$~0.03 & 0.18~$\pm$~0.03 & 0.38~$\pm$~0.03 &	0.19~$\pm$~0.03 & 0.16~$\pm$~0.03 & -3.07~$\pm$~0.12 & 2.24~$\pm$~0.13 & 4 \\
\smallskip
(2,8)	& 0.10~$\pm$~0.04 & 0.10~$\pm$~0.04 & 0.25~$\pm$~0.04 &	0.07~$\pm$~0.04 & 0.10~$\pm$~0.04 & -3.42~$\pm$~0.12 & 4.10~$\pm$~0.12 & 4 \\
\smallskip
(4,-2)	& 0.04~$\pm$~0.03 & 0.07~$\pm$~0.03 & 0.19~$\pm$~0.03 &	0.05~$\pm$~0.03 & 0.08~$\pm$~0.03 & -4.85~$\pm$~0.13 & 3.95~$\pm$~0.14 & 4 \\
\smallskip
(4,0)	& 0.15~$\pm$~0.03 & 0.18~$\pm$~0.03 & 0.47~$\pm$~0.03 &	0.14~$\pm$~0.03 & 0.12~$\pm$~0.03 & -4.75~$\pm$~0.12 & 4.09~$\pm$~0.13 & 8 \\
\smallskip
(4,2)	& 0.10~$\pm$~0.03 & 0.14~$\pm$~0.03 & 0.35~$\pm$~0.03 &	0.12~$\pm$~0.03 & 0.13~$\pm$~0.03 & -4.26~$\pm$~0.12 & 4.23~$\pm$~0.12 & 7 \\
         \hline
      \end{tabular}
      \begin{list}{}{}
         \item[$^{\mathrm{a}}$] Offset in relation to the equatorial coordinates $\alpha(1950) = 17\rm^{h} 17\rm^{m} 32\rm^{s}$ 
            and $\delta(1950) = -35\degr 42\arcmin$.
         \item[$^{\mathrm{b}}$] Velocity of the main line.
         \item[$^{\mathrm{c}}$] Number of observations made in each position.
      \end{list}
   \end{table*}
%
%
%

   \begin{figure*}
      {\includegraphics{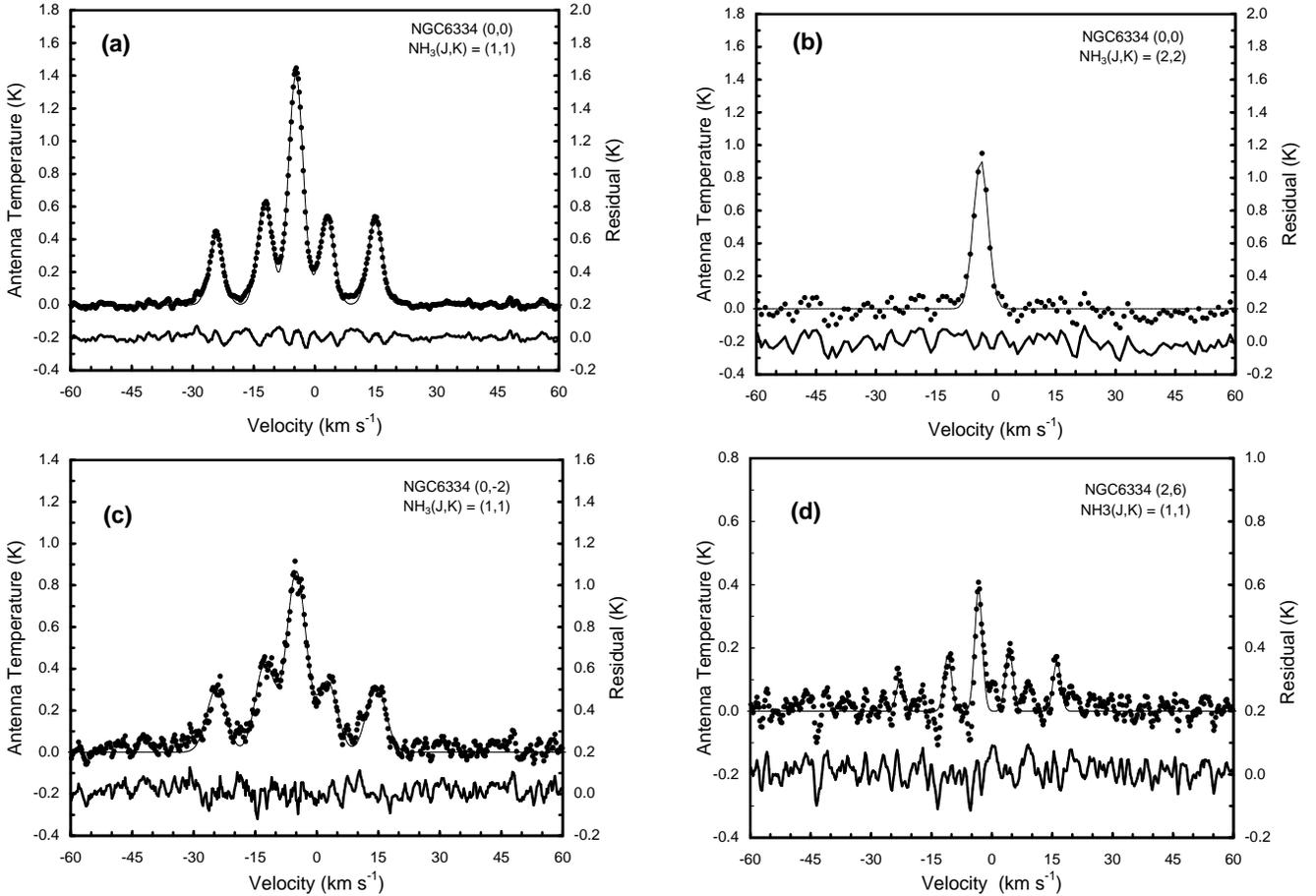}}
      \caption[]{Spectra from the central ($\Delta\alpha,\Delta\delta$) = (0\arcmin,0\arcmin) position for the 
                 transitions {\bf (a)} NH$_3$(J,K) = (1,1) and {\bf (b)} (2,2). The NH$_3$(J,K) = (1,1) spectrum 
                 from the {\bf (c)} (0\arcmin,-2\arcmin) and {\bf (d)} (2\arcmin,6\arcmin) positions are 
                 also shown. The thin lines represent the Gaussian fittings, while the lines, below the observed 
                 spectra, are the residuals.}
         \label{FigSpect}
   \end{figure*}
%

   The NH$_3$(J,K) = (1,1) spectra in all the observed positions exhibit systematic differences between 
   the line intensities in each pair of hyperfine transitions (F=1$\rightarrow$0, F=0$\rightarrow$1) and 
   (F=1$\rightarrow$2, F=2$\rightarrow$1), which are, at least in the strongest points, much larger than 
   the rms measured at the baseline. These intensity differences can be interpreted as the signature of 
   non-LTE conditions. However, we should point out that this conclusion cannot be obtained by looking 
   only at the residuals of the line fittings, which present larger rms than the baseline, both under LTE 
   and non-LTE assumption. This is expected an expected result, since the superposition of sources and the 
   existence of velocity gradients can make the line shapes differ from Gaussian functions. 

   The antenna temperature distribution and gradients of velocity and line-width are presented in Figure 2. 
   They agree with the results of Forster et al. (\cite{for87}) and Kuiper et al. (\cite{kui95}), who mapped 
   with high spatial resolution a small region (about 100\arcsec\ in diameter) around the central position in 
   our map in the NH$_3$(J,K) = (3,3) transition. However, the distribution of velocities differs from that 
   found by Schwartz et al. (\cite{sch78}), since their map do not show an increase in the velocity's modulus 
   towards negative declinations, possibly due to the low signal-to-noise ratio in their spectra.

   \begin{figure*}
      {\includegraphics{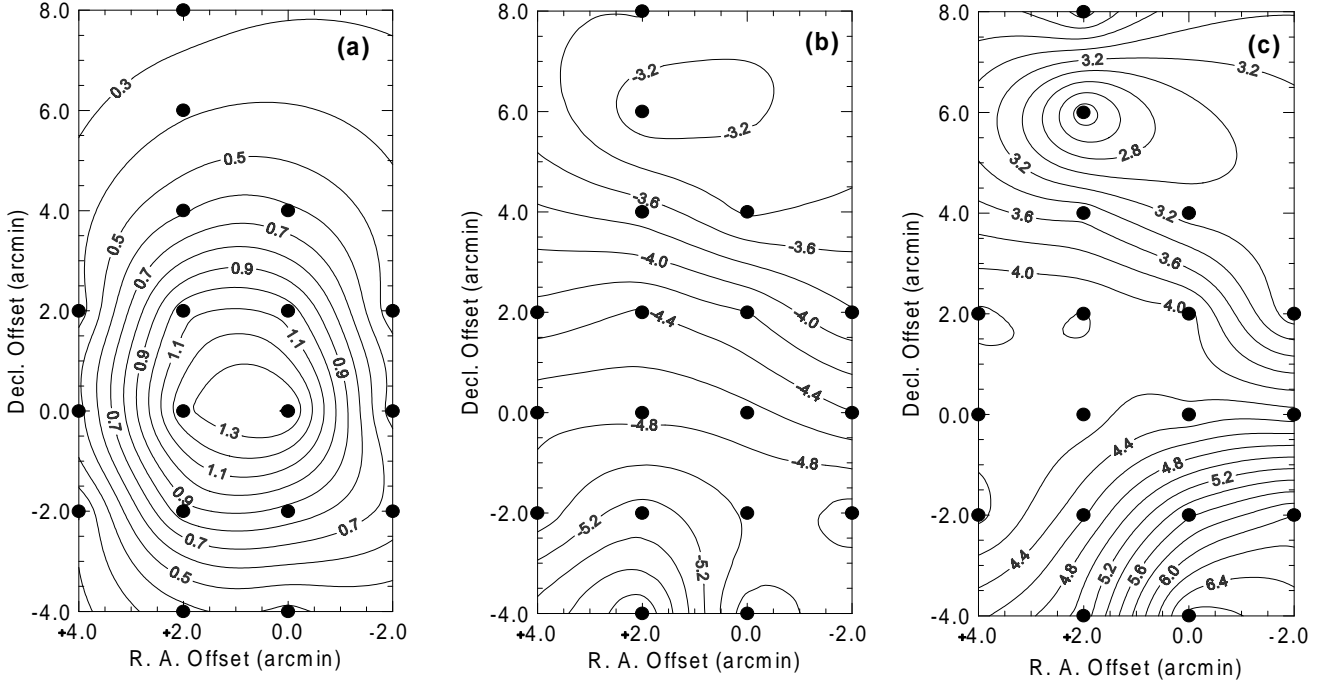}}
      \caption[]{Maps of {\bf (a)} antenna temperature (in units of K), {\bf (b)} velocity 
                 (in units of km s$^{-1}$) and {\bf (c)} line-width (in units of km s$^{-1}$) 
                 for the NGC 6334 region. The (0\arcmin,0\arcmin) position has coordinates 
                 $\alpha(1950) = 17^{h} 17^{m} 32^{s}$ and $\delta(1950) = -35\degr 42\arcmin$. 
                 The black dots mark the observed positions.
                 }
         \label{FigMap}
   \end{figure*}
%
 
%

\section{Results}

\subsection{Identification of the sources NGC\,6334\,I, NGC\,6334\,I(N)w and NGC\,6334\,I(N)e}

   Although the map in Figure 2 seems to be consistent with the presence of a single source 
   centered in NGC\,6334\,I(N), high resolution observations (JHH88; Kraemer \& Jackson \cite{krja99}) 
   showed the existence of at least two ammonia features, with angular sizes of the order and smaller 
   than 1\arcmin, centered at the positions of the continuum sources NGC\,6334\,I(N) and NGC\,6334\,I 
   respectively. Taking into account the half-power beam width of the Itapetinga radio telescope, we 
   find that the expected contribution of these two sources at the position ($\Delta\alpha,\Delta\delta$) 
   = (2\arcmin,6\arcmin) [hereafter, just (2\arcmin,6\arcmin)] should fall below the detection limit given 
   by the rms of the observations. Since this is not the case, as can be seen in Figure 1(d), we conclude 
   that another source is present in the region. We will describe, in the following paragraphs, the procedure 
   used to separate the contribution of each component.

   NGC\,6334\,I, when observed with high resolution (JHH88), exihibts two main structures, with velocities 
   centered in -6.6 and -9.4 km s$^{-1}$ and widths of 2.0 and 3.5 km s$^{-1}$ respectively. The maximum flux 
   densities of these structures are 0.42 and 0.48 Jy beam$^{-1}$ respectively which, once diluted into the 
   4.2\arcmin\ Itapetinga radio telescope beam, correspond to antenna temperatures of 0.20 and 0.23 K respectively. 
   We calculated the contribution of this source to the NH$_3$(J,K) = (1,1) spectra in our map as the convolution 
   of the antenna beam and two point sources, with the characteristics mentioned above. The intensity of the 
   hyperfine satellite lines was calculated assuming LTE in an optically thin source. Although the antenna 
   temperature of NGC\,6334\,I is small compared to the intensity of NGC\,6334\,I(N), its effect on the line-width 
   is significant, and its subtraction decreased the observed gradient along the map.

   Since the high resolution observations of Kraemer \& Jackson (\cite{krja99}) have revealed NGC\,6334\,I(N) as 
   an extended source, although smaller than the Itapetinga radio telescope beam, it was difficult to subtract 
   its contribution to expose the third and much weaker source. For this reason we used the spectrum at position 
   (2\arcmin,6\arcmin), where we did not expect any contribution from the strongest source, as representative of 
   the spectrum of the weakest. We labeled these sources as NGC\,6334\,I(N)w and NGC\,6334\,I(N)e [hereafter, 
   I(N)w and I(N)e] respectively. We assumed no velocity gradients and subtracted a fraction of the (2\arcmin,6\arcmin) 
   spectrum from all the others. This fraction was chosen in such a way that the resultanting spectra had the 
   same departure from LTE as found at position (0\arcmin,0\arcmin). Moreover, since there is a velocity gradient 
   along the region, the subtracted fraction was also limited in order to avoid fake absorptions in the resulting 
   spectra. 

   \begin{figure}
      {\includegraphics{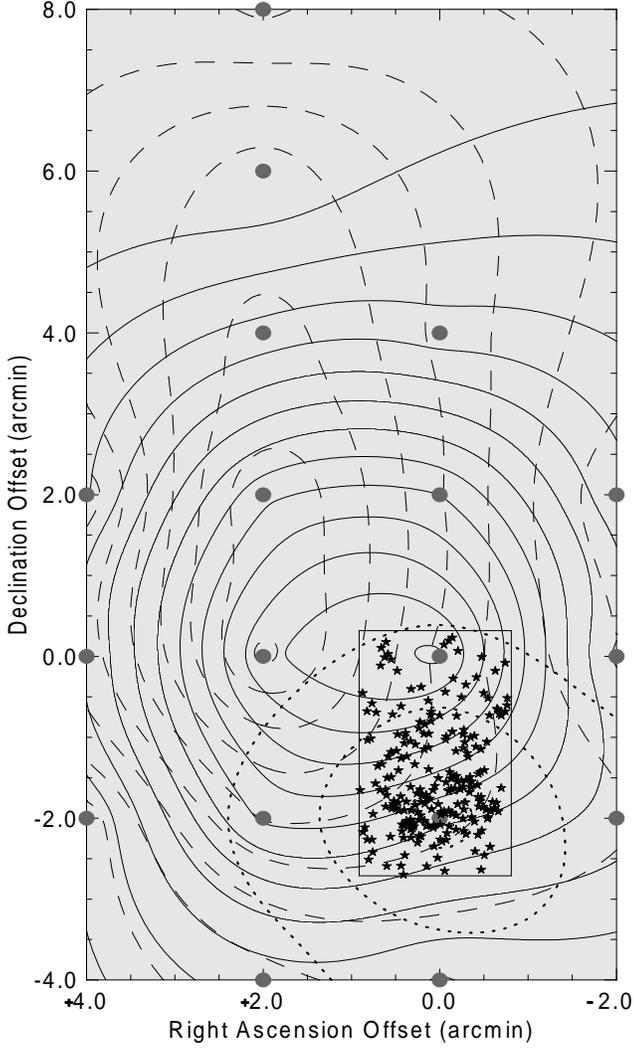}}
      \caption[]{The superposition of the contribution of the three sources: NGC\,6334\,I (dotted 
                 lines), I(N)w (solid lines) and I(N)e (dashed lines). The contours are given in 
                 antenna temperature, starting at 0.1 and ending at 1.3 K (with spacing of 0.1 K). 
                 The gray points mark the observed positions. The rectangle shows the area that 
                 was observed by Tapia et al. \cite{tap96} in the JHK bands. The black stars are 
                 the sources which are detected by these authors.
                 }
         \label{FigSuperp}
   \end{figure}
%

   In Figure 3, we show the superimposed antenna temperature maps of the three regions, NGC\,6334\,I, I(N)w and I(N)e. 
   We can see that I(N)w is aproximately 2.5 and 6.5 times brighter than I(N)e and NGC\,6334\,I, respectively. The 
   source I(N)w turned out to be smaller than the beam size, as expected from the observations of Kraemer \& Jackson 
   (\cite{krja99}). On the other hand, the I(N)e region looks more elongated in declination. 

   The subtraction procedures affected the magnitude of the gradients of velocity and line-width, 
   but not their directions. In fact, the gradients became softer than before, showing that the other 
   sources were responsible in some amount for the variations of these properties along the region 
   I(N)w. As it can be seen in Figure 4, the velocity gradient is aproximately perpendicular to the 
   Galactic plane. According to the model adopted here, the gradient in line-width is related to gradient 
   in turbulence, which turned out to be larger in the direction of the more evolved sources.  

   \begin{figure}
      {\includegraphics{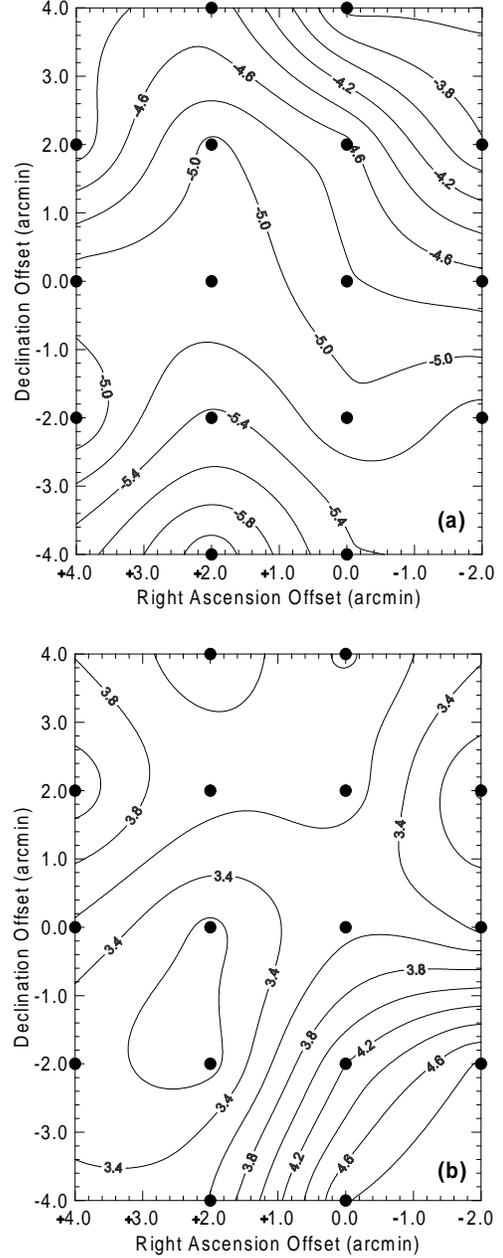}}
      \caption[]{The gradients of {\bf (a)} velocity and {\bf (b)} line-width for region I(N)w. The 
                 black points mark the observed positions. The countour values are given in units of 
                 km s$^{-1}$.
                }
         \label{FigGradINa}
   \end{figure}
%

\subsection{The physical parameters of I(N)w and I(N)e regions in non-LTE conditions}

   We used the antenna temperature of the hyperfine satellites relative to the main line in the 
   NH$_3$(J,K) = (1,1) spectrum and the ratio between the antenna temperatures of the main lines 
   of NH$_3$(J,K) = (2,2) and (1,1), shown in Table 2, to calculate the relevant physical parameters 
   for the I(N)w and I(N)e regions. We applied the model of SW85 to obtain the NH$_3$ column density 
   ($N_{\mathrm{NH_3}}$), the kinetic temperature ($T_{\mathrm{K}}$) and the H$_2$ density 
   ($n_{\mathrm{H_2}}$) of the clumps in the two regions. In region I(N)w we used the spectrum from 
   the (0\arcmin,0\arcmin) position, which corresponds to the highest intensity, and in I(N)e we used 
   the (2\arcmin,6\arcmin) position, where we believe there is not important contribution from I(N)w. 

   To determine the physical parameters of the clumps we looked into the ($N_{\mathrm{NH_3}}$, $n_{\mathrm{H_2}}$, 
   $T_{\mathrm{K}}$) space for the point at which, according to the model of SW85, the curves representing 
   the observed line intensity ratios intercept themselves. We had to extrapolate these curves up to 
   H$_2$ densities of 10$^8$ cm$^{-3}$ to obtain a good crossing point, which occured at $n_{\mathrm{H_2}} 
   \sim 10^{7.5}$ cm$^{-3}$. 

   In Figure 5, we show the curves of constant ratio between the hyperfine satellites and the main line 
   in the NH$_3$(J,K) = (1,1) spectrum, as well as the ratio between the main line intensities of the NH$_3$(J,K) 
   = (2,2) and (1,1) transitions for a kinetic temperature of 24 K, which corresponds to the best fit in both 
   regions. In Table 3, we present the values of the physical parameters extracted from this figure: the brightness 
   temperature for the NH$_3$(J,K) = (1,1) and (2,2) transitions [$T_{\mathrm{B}}(1,1)$ and $T_{\mathrm{B}}(2,2)$, 
   respectively], the kinetic temperature, the NH$_3$ column density and the H$_2$ density. We can see that the 
   physical conditions of the clumps in both regions are pratically the same. It is important to stress that the 
   values listed in Table 3 would not change substantially if we use the original spectra, including the 
   contribution of the three sources, since I(N)w is much brighter than the others. On the other hand, the 
   dispersion of the contours around the intersection point in Figure 5(a) decreased appreciably after the 
   separation of the three components, resulting in a better accuracy in the determination of the physical 
   parameters of the region.

   \begin{table}
      \caption[]{Ratio between the observed antenna temperatures of each hyperfine satellite in relation to the 
                 (1,1) main line [$T_{\mathrm{A}}^m$], besides the ratio between the antenna temperature of 
                 the (2,2) and (1,1) main lines.}
      \label{KapSou}
      \begin{tabular}{lcc}
         \hline \smallskip
           & NGC\,6334\,I(N)w & NGC\,6334\,I(N)e \\
         \hline
\smallskip
$T_{\mathrm{A}}$(F=1$\rightarrow$0) / $T_{\mathrm{A}}^m$ & 0.295~$\pm$~0.008 & 0.34~$\pm$~0.09 \\
\smallskip
$T_{\mathrm{A}}$(F=1$\rightarrow$2) / $T_{\mathrm{A}}^m$ & 0.409~$\pm$~0.008 & 0.47~$\pm$~0.09 \\
\smallskip
$T_{\mathrm{A}}$(F=2$\rightarrow$1) / $T_{\mathrm{A}}^m$ & 0.371~$\pm$~0.008 & 0.50~$\pm$~0.09 \\
\smallskip
$T_{\mathrm{A}}$(F=0$\rightarrow$1) / $T_{\mathrm{A}}^m$ & 0.364~$\pm$~0.008 & 0.42~$\pm$~0.09 \\
\smallskip
$T_{\mathrm{A}}$(2,2) / $T_{\mathrm{A}}^m$	         &  0.67~$\pm$~0.04  &         -       \\
         \hline
      \end{tabular} 
   \end{table}
%
%

\subsection{Physical parameters of the individual clumps}

   As it was previously mentioned, to explain the hyperfine structure intensity anomalies in the
   NH$_3$(J,K) = (1,1) transition by the superposition of individual clump spectra, their dispersion 
   velocities must be small, implying that they should be gravitationally stable. Therefore, their 
   individual masses ($M_{\mathrm{cl}}$) cannot be larger than the Jeans mass ($M_{\mathrm{J}}$) and 
   their diameters ($L_{\mathrm{cl}}$) must be smaller or equal than the Jeans length ($L_{\mathrm{J}}$). 
   Arguments in favor of these hypotheses have been presented by Curry \& McKee (\cite{cumc00}) based on 
   numerical simulations. Assuming spherical clumps with a ratio of helium to molecular hydrogen of 
   0.25, we calculated their diameter and mass, using the following expressions:
   \begin{eqnarray}
      L_{\mathrm{cl}} = L_{\mathrm{J}} = \sqrt{ \frac{ \pi\, k\, T_{\mathrm{K}}}{7.2\, G\, m_{\mathrm{H}}^2\, n_{\mathrm{H_2}}}}\
   \end{eqnarray}
   \begin{eqnarray}
      M_{\mathrm{cl}} = M_{\mathrm{J}} = \frac{ \pi}{2} m_{\mathrm{H}}\, n_{\mathrm{H_2}}\, L_{\mathrm{cl}}^3\
   \end{eqnarray}
   where $k$ is the Boltzmann's constant, $m_{\mathrm{H}}$ the mass of hydrogen atom and $G$ the 
   gravitational constant.

   \begin{figure}
      {\includegraphics{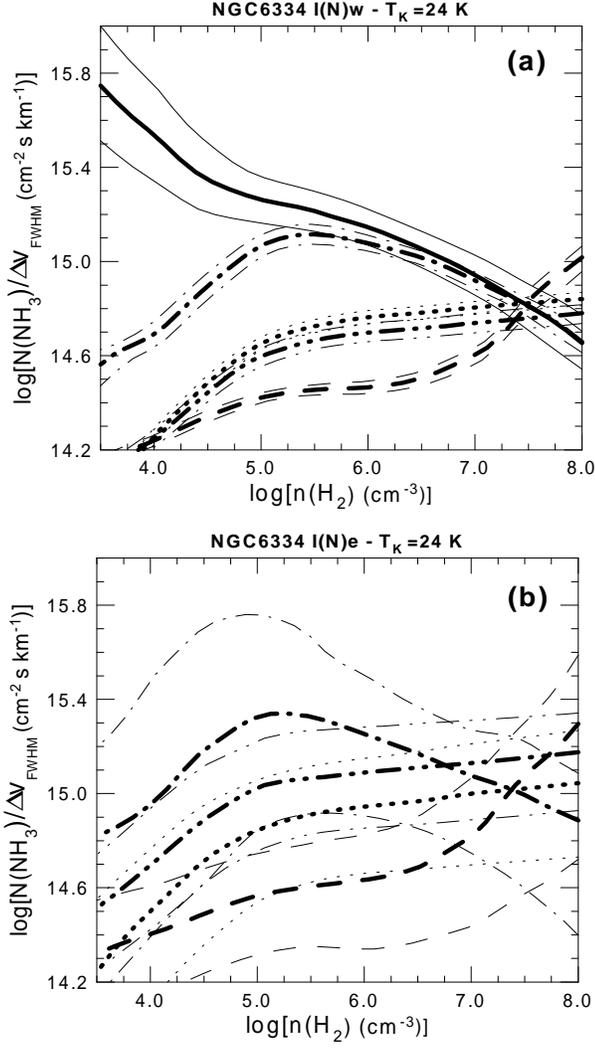}}
      \caption[]{Ratio between the intensity of the satellites and the main line of the NH$_3$(J,K) = 
                 (1,1) spectrum, represented by: dashed, dashed dotted, dashed dotted dotted and dotted 
                 lines (corresponding to F=1$\rightarrow$0, F=1$\rightarrow$2, F=2$\rightarrow$1 and 
                 F=0$\rightarrow$1 transitions, respectively) for a kinetic temperature of 24 K. The 
                 upper and lower plots [(a) and (b) respectively] are associated to the I(N)w and I(N)e 
                 in this order. Ratio between the main lines intensity of the NH$_3$(J,K) = (2,2) and (1,1) transitions 
                 is shown by the solid lines. The weaker lines are the uncertainties associated to the 
                 fitting. The locus where these curves intersect each other represents the physical 
                 parameters of the region. 
                }
         \label{FigStutINa}
   \end{figure}
%

   The beam filling factor ($f_{\mathrm{JK}}$) can be calculated from the brightness temperature, $T_{\mathrm{B}}(J,K)$, 
   derived from the model and $T_{\mathrm{B}}^{obs}(J,K)$, obtained from the observations, which is proportional to the 
   antenna temperature:
   \begin{eqnarray}
      f_{\mathrm{JK}} = \frac{T_{\mathrm{B}}^{obs}(J,K)}{T_{\mathrm{B}}(J,K)} = \frac{2\,c^2}{\pi\, k\, \nu^2} \frac{A}{\theta_{\mathrm{B}}^2} \frac{T_{\mathrm{A}}(J,K)}{T_{\mathrm{B}}(J,K)}\
   \end{eqnarray}
   where $\theta_{\mathrm{B}}$ is the half-power beam width of the telescope and $A$ is the ratio between 
   flux density and antenna temperature of a continuum point source. In this work, we used the radio galaxy 
   Virgo A as the calibrating source.

   Using the definition of beam filling factor, we can determine the number 
   $K_{\mathrm{cl}}$ of clumps inside the beam's solid angle:
   \begin{eqnarray}
      K_{\mathrm{cl}} = f_{\mathrm{JK}} \left( \frac{D\, \theta_{\mathrm{B}}}{L_{\mathrm{cl}}} \right)^{2}\
   \end{eqnarray}
   where $D$ is the distance from the cloud.

   To obtain the density of clumps ($\zeta_{\mathrm{cl}}$), we can consider two different situations. The 
   first one corresponds to the clumps spread over a solid angle larger than the beam. In this case, the 
   density of clumps is given by:
   \begin{eqnarray}
      \zeta_{\mathrm{cl}} = \frac{4}{\pi} \frac{K_{\mathrm{cl}}}{D^3\, \theta_{\mathrm{B}}^2 \sqrt{\theta_{\mathrm{\alpha}}\, \theta_{\mathrm{\beta}}}}\
   \end{eqnarray}
   where $\theta_{\mathrm{\alpha}}$ and $\theta_{\mathrm{\beta}}$ are, respectively, the extension of the region 
   in right ascension and declination. In the second one, if the source is smaller than the antenna beam, we can 
   only estimate either an upper limit, considering that the clumps are grouped side by side and without superposition, 
   or a lower limit, supposing they are spread over the whole beam:
   \begin{eqnarray}
   \frac{6}{\pi} \frac{K_{\mathrm{cl}}}{\left( D\, \theta_{\mathrm{B}} \right)^3} \leq \zeta_{\mathrm{cl}} \leq \frac{6}{\pi} \frac{K_{\mathrm{cl}}}{\left( D\, \theta_{\mathrm{B}}\, \sqrt{f_{\mathrm{JK}}} \right)^3}\
   \end{eqnarray}

   \begin{table}
      \caption[]{The physical parameters determined from the Figure 5 for I(N)w and I(N)e regions.}
      \label{KapSou}
      \begin{tabular}{lcc}
         \hline \smallskip
           & NGC\,6334\,I(N)w & NGC\,6334\,I(N)e \\
         \hline
\smallskip
$T_{\mathrm{B}}$(1,1) (K)                                  & 12.6~$\pm$~0.4 & 15.5~$\pm$~1.0 \\
\smallskip
$T_{\mathrm{B}}$(2,2) (K)                                  &  8.4~$\pm$~0.6 & 11.6~$\pm$~1.1 \\
\smallskip
$T_{\mathrm{K}}$ (K)$^{\mathrm{a}}$                        & 24.0~$\pm$~0.8 & 24.0~$\pm$~1.0 \\
\smallskip
$N_{\mathrm{NH_3}}$ (10$^{15}$ cm$^{-2}$)                  &  2.3~$\pm$~0.8 &  2.4~$\pm$~0.8 \\
\smallskip
$n_{\mathrm{H_2}}$ (10$^{7}$ cm$^{-3}$)$^{\mathrm{a,b}}$   &  1.8~$\pm$~0.4 &  1.4~$\pm$~0.7 \\
         \hline
      \end{tabular}
      \begin{list}{}{}
         \item[$^{\mathrm{a}}$] For each clump.
         \item[$^{\mathrm{b}}$] According to SW85, the H$_2$ density derived from their model must be 
                                divided by 1.75 in order to obtain the true value for this parameter, shown 
                                in this table. The reason for this procedure is that only NH$_3$ - He 
                                collisions were considered in the numerical model of SW85 as excitational 
                                mechanism of the ammonia molecule. 

      \end{list}
   \end{table}
%
%

   Once the density of clumps and their diameters are known, we can calculate their collision time 
   ($t_{\mathrm{coll}}$):
   \begin{eqnarray}
      t_{\mathrm{coll}} = \frac{8\, \sqrt{ln2}}{ \pi\, L_{\mathrm{cl}}^2\, \zeta_{\mathrm{cl}}\, \Delta v_{\mathrm{FWHM}}^{JK}} 
   \end{eqnarray}

   We can also calculate the mean H$_2$ density ($\bar n_{\mathrm{H_2}}$), which is the density of an homogeneous 
   cloud with mass equal to the total mass of the clumps filling the telescope beam:
   \begin{eqnarray}
      \bar n_{\mathrm{H_2}} = \frac{M_{\mathrm{cl}}}{3\, m_{\mathrm{H}}} \zeta_{\mathrm{cl}} 
   \end{eqnarray}

   The ammonia abundance is calculated from the following expression:
   \begin{eqnarray}
      X_{\mathrm{NH_3}} = \frac{N_{\mathrm{NH_3}}}{L_{\mathrm{cl}}\,n_{\mathrm{H_2}}}
   \end{eqnarray}
   where it was assumed that there is not superposition between the clumps, so that the line 
   of sight extension is equal to the diameter of each clump (SW85).

   The physical parameters of the clumps in regions I(N)w and I(N)e and their respective 
   uncertainties are listed in Table 4. The quoted errors for the mass and diameter of the clumps 
   represent the uncertainties in the kinetic temperature and H$_2$ density. 

   The clumps in both regions have similar diameters (0.007 pc), masses (0.2 M$_{\sun}$) and ammonia abundances 
   (7$\times$10$^{-9}$). They also seem to be gravitationally bounded, since the total virial mass is about a 
   factor of ten lower than the total mass of the clumps which are contained in the beam's solid angle [7600 and 
   2100 M$_{\sun}$ for I(N)w and I(N)e respectively]. The 36000 clumps of I(N)w are spread in such a way that their 
   density is between 6000 and 33000 clumps per pc$^{-3}$, much larger than the 720 clumps per pc$^{-3}$ in the 
   I(N)e region. 

   The beam filling factor for the I(N)w region is larger than expected from the NH$_3$(J,K) = (3,3) data 
   presented in Kraemer \& Jackson (\cite{krja99}) (around 0.10 for a beam size of 4\arcmin). The reason is that 
   we had assumed no superposition between the clumps along the line of sight. The physical superposition required 
   by the smaller filling factor is small enough to guarantee very little velocity superposition, in which case the 
   spectra of the superposed clumps can be simply added. 

   \begin{table}
      \caption[]{The physical parameters for I(N)w and I(N)e regions calculated in non-LTE conditions. The 
                 uncertainties are given in parenthesis.}
      \label{KapSou}
      \scriptsize
      \begin{tabular}{lcc}
         \hline \smallskip
           & NGC\,6334\,I(N)w & NGC\,6334\,I(N)e \\
         \hline
\smallskip
$L_{\mathrm{cl}}$ ($10^{-3}$ pc)                           &   6.7(0.8)                                         &  7.5(1.7)    \\
\smallskip
$M_{\mathrm{cl}}$ ($M_{\sun}$)                             &  0.21(0.03)                                        & 0.24(0.06)   \\
\smallskip
$f_{\mathrm{JK}}$                                          &  0.33(0.03)                                        & 0.10(0.01)   \\
\smallskip
$K_{\mathrm{cl}}$ ($10^{3}$)                               &    36(10)                                          &    9(4)      \\
\smallskip
$\zeta_{\mathrm{cl}}$ (10$^{3}$ pc$^{-3}$)                 & 6(2)~$\leq\zeta_{\mathrm{cl}}\leq$~33(11)          & 0.72(0.36)   \\
\smallskip
$t_{\mathrm{coll}}$ (10$^{5}$ years)                       & 4(2)~$\leq t_{\mathrm{coll}}\leq$~22(9)            &  230(155)    \\
\smallskip
$\bar n_{\mathrm{H_2}}$ (10$^{4}$ cm$^{-3}$)               & 1.7(0.6)~$\leq\bar n_{\mathrm{H_2}}\leq$~9.5(3.4)  & 0.23(0.13)   \\
\smallskip
$X_{\mathrm{NH_3}}$ (10$^{-9}$)                            &   6.2(2.3)                                         &  7.2(3.0)    \\
\smallskip
$M_{\mathrm{tot}} (10^{3} M_{\sun})^{\mathrm{a}}$          &   7.6(2.3)                                         &  2.1(1.2)    \\
\smallskip
$M_{\mathrm{vir}} (10^{3} M_{\sun})^{\mathrm{b}}$          &  0.67(0.06)                                        & 0.15(0.02)   \\
         \hline
      \end{tabular} 
      \begin{list}{}{}
         \item[$^{\mathrm{a}}$] Total mass of whole clumps which are contained in the telescope beam.
         \item[$^{\mathrm{b}}$] Mass obtained from the Virial Theorem.
       \end{list}
   \end{table}
%
%

\section{Discussion}

\subsection{Ammonia abundance for the regions I(N)w and I(N)e}

   The study of ammonia abundance is important because it provides observational constrains for the chemical 
   models that try to explain how this molecule is formed in the interstellar medium, and it may be used 
   as an age indicator of the molecular cloud (e.g., Suzuki et al. \cite{suz92}). Although the chemical 
   models calculate the evolution of the ammonia abundance as a function of time with a limited confidence, 
   due mainly to the uncertainties concerning the formation of this molecule in the interstellar medium, 
   they agree that the time necessary to reach the equilibrium abundance is approximately 10$^6$ years. If 
   depletion effects are not consider in the model, the equilibrium abundance for the ammonia molecule is 
   10$^{-7}$. If there is depletion, this value might be 10$^{-8}$ (Bergin et al. \cite{ber95}) or 10$^{-9}$ 
   (Hasegawa \& Herbst \cite{hahe93}), even though in this last case the abundance reaches equilibrium after 
   10$^8$ years.

   Comparing the ammonia abundances listed in Table 4 with the equilibrium abundance, we conclude that either 
   the regions I(N)w and I(N)e have an age lower than 10$^6$ years, or that ammonia depletion is important in 
   both sources. This result was also obtained by Kuiper et al. (\cite{kui95}), assuming LTE conditions. Another 
   relevant aspect that must be taken into account is the dependence of the equilibrium abundance on the nitrogen 
   abundance in the interstellar medium, and therefore the gradient of nitrogen with galactocentric distance must 
   be also considered (Vilas-Boas \& Abraham \cite{viab00}).

\subsection{Comparison with LTE models}

   In this sub-section, we will derive, under assumptions of LTE, the physical parameters of the source I(N)w 
   and compare them with the parameters derived in section 3 under non-LTE conditions. We calculate these 
   parameters at the (0\arcmin,0\arcmin) position because it is the only point in which the NH$_3$(J,K) = (2,2) 
   transition was observed. In the LTE calculation we considered a three-level system, in order to obtain the 
   rotational and kinetic temperatures ($T_{\mathrm{R}}$ and $T_{\mathrm{K}}$ respectively) through the equations 
   (Ho \& Townes \cite{hoto83}; Walmsley \& Ungerechts \cite{waun83}):
   \begin{eqnarray}
      T_{\mathrm{R}} = -41.5\times \left[\ln\left(\frac{9}{20}\frac{\tau_{22}}{\tau_{11}}\frac{\Delta v_{\mathrm{FWHM}}^{22}}{\Delta v_{\mathrm{FWHM}}^{11}}\right)\right]^{-1}
   \end{eqnarray}
   \begin{eqnarray}
      T_{\mathrm{K}} = \frac{T_{\mathrm{K}}}{1+\left(T_{\mathrm{K}}/41.7~K\right)\ln\left[1+\frac{C\left(22\rightarrow21\right)}{C\left(22\rightarrow11\right)}\right]}
   \end{eqnarray}
   where $\tau_{11}$ and $\tau_{22}$ are respectively the optical depths in the transitions NH$_3$(J,K) = (1,1) 
   and (2,2), C(22 $\rightarrow$ 21) and C(22 $\rightarrow$ 11) are the collision rate coefficients given 
   by Danby et al. (\cite{dan88}).

   To determine the optical depth, we assumed that the cloud has a spherical shape, that the excitation temperature 
   ($T_{\mathrm{ex}}$) is the same in all hyperfine lines of the NH$_3$(J,K) = (1,1) spectrum (Ho \& Townes 
   \cite{hoto83}). Thus, using the ratio between the (1,1) main line and its inner hyperfine satellites, $\tau_{11}$ 
   can be calculated from:
   \begin{eqnarray}
      \frac{T_{\mathrm{A}}^{\mathrm{m}}(1,1)}{T_{\mathrm{A}}^{\mathrm{s}}(1,1)} = \frac{1-e\left(\tau_{11}^{\mathrm{m}}\right)}{1-e\left(\tau_{11}^{\mathrm{s}}\right)}
   \end{eqnarray}
   where $\tau_{11}^{\mathrm{m}}$ and $\tau_{11}^{\mathrm{s}}$ are, respectively, the optical depth for the 
   main line and the inner satellites, with $\tau_{11}^{\mathrm{s}}$ = 0.28 $\tau_{11}^{\mathrm{m}}$ (Ho \& 
   Townes \cite{hoto83}) and $e(\tau) = 2\,\tau^{-2}[1-e^{-\tau}(1+\tau)]$ (SW85). To obtain the optical depth 
   for the NH$_3$(J,K) = (2,2) transition, it is necessary to substitute in (12) $T_{\mathrm{A}}^{\mathrm{s}}(1,1)$ 
   by $T_{\mathrm{A}}(2,2)$, the antenna temperature of the (2,2) main line.

   As we have less equations than unknown parameters, it is necessary to make some assumptions to solve the 
   system. There are at least two clear alternatives: we may assume that either the excitation and rotational 
   temperatures have the same value, or that the source fills completely the antenna beam. Here, we adopted the 
   first alternative since high resolution observations indicate filling factors smaller than one for these 
   sources (Kuiper et al. \cite{kui95}; Kraemer \& Jackson \cite{krja99}; Megeath \& Tieftrunk \cite{meti99}).

   We present in Table 5 the physical parameters obtained under LTE and non-LTE conditions. The NH$_3$ column 
   density and the ammonia abundance are not sensitive to LTE departures. The largest discrepancies are found 
   in the kinetic and rotational temperatures. In fact, the kinetic temperature estimated under LTE conditions is 
   about twice the value obtained in non-LTE. The optical depth of I(N)w, in both transitions, is slightly 
   smaller than what it was found in previous studies (e.g., Forster et al. \cite{for87}; Vilas-Boas et al. \cite{vil88}; 
   Kuiper et al. \cite{kui95}). We believe that the differences can be attributed to the lower signal-to-noise 
   ratio in previous observations. The beam filling factor is larger under LTE conditions. This was already 
   expected because under non-LTE conditions the gas is confined in clumps and not spread in a homogeneous cloud.

\subsection{Evolutionary state of the regions}

   We will try to determine now the evolutionary state of the three regions identified in our ammonia 
   observations. A possible way to explore this question is comparing the distribution of indicators of 
   early stages of star formation. In Figure 3, we show the distribution of infrared sources, detected 
   in the JHK bands by Tapia et al. (\cite{tap96}) in a small region of the mapped area. We can see a 
   decrease in the number of infrared sources towards the region I(N)w and also the existence of a 
   cluster of sources around NGC\,6334\,I, indicating that it is more evolved than I(N)w. Other 
   observational results favorable to this scenario are the detection of an UC\ion{H}{ii} region and 
   the identification of a molecular bipolar outflow associated to this source (Rodr\'\i guez et al. 
   \cite{rod82}; Bachiller \& Cernicharo \cite{bace90}). There are no infrared observations towards 
   I(N)e, to indicate its evolutionary stage. 

   Our observations, together with the non-LTE model (SW85), show that the physical properties of the 
   clumps in I(N)w and I(N)e are similar. However, the volume density of clumps in I(N)w ($\geq$ 6000 pc$^{-3}$) 
   is at least a hundred times higher than in I(N)e, resulting in a collision rate which is hundred times 
   higher in I(N)w than I(N)e. Thus, assuming that collisions result in coagulation of the fragments, which 
   finally collapse to form stars after several collision times (SW85; Scalo \& Pumphrey \cite{scpu82}), 
   we expect that I(N)e is in a earlier star formation stage than I(N)w.

   The evolution scenario proposed by SW85 and the results presented in this work indicate that the regions 
   I(N)w and I(N)e could form in the future star clusters, similar to those detected in NGC\,6334\,I. 

   \begin{table}
      \caption[]{Comparison between the physical parameters for I(N)w, calculated in LTE and non-LTE 
                 conditions.}
      \label{KapSou}
      \scriptsize
      \begin{tabular}{lcc}
         \hline \smallskip
           & LTE & non-LTE \\
         \hline
\smallskip
$T_{\mathrm{K}}$ (K)                           &   44~$\pm$~8      & 24.0~$\pm$~0.8   \\
\smallskip
$T_{\mathrm{R}}$ (K)                           &   28~$\pm$~3      & 18.7~$\pm$~0.3   \\
\smallskip
$\tau_{\mathrm{11}}$                           & 1.42~$\pm$~0.01   &        -         \\
\smallskip
$\tau_{\mathrm{22}}$                           & 0.71~$\pm$~0.08   &        -         \\
\smallskip
$f_{\mathrm{11}}$                              & 0.88~$\pm$~0.10   & 0.33~$\pm$~0.03  \\
\smallskip
$f_{\mathrm{22}}$                              & 0.88~$\pm$~0.10   & 0.33~$\pm$~0.04  \\
\smallskip
$\bar n_{\mathrm{H_2}}$ (10$^{4}$ cm$^{-3}$)   &   12~$\pm$~5      &  1.7~$\pm$~0.6~$\leq\bar n_{\mathrm{H_2}}\leq$~9.5~$\pm$~3.4 \\
\smallskip
$N_{\mathrm{NH_3}}$ (10$^{15}$ cm$^{-2}$)      &  2.5~$\pm$~0.1    &  2.3~$\pm$~0.8    \\
\smallskip
$X_{\mathrm{NH_3}}$ (10$^{-9}$)                &  3.3~$\pm$~1.7    &  6.2~$\pm$~2.3    \\
         \hline
      \end{tabular} 
   \end{table}
%
%

\section{Conclusions}

   The NH$_3$(J,K) = (1,1) spectra from the NGC 6334 region exhibit hyperfine structure intensity anomalies, 
   suggesting departures from LTE conditions. We have used the numerical model of SW85, based on the 
   qualitative model of Matsakis et al. (\cite{mat77}) to calculate the physical parameters of the ammonia 
   clouds. This model takes into account the non-LTE effects through selective trapping in the far IR 
   transition NH$_{3}$(J,K) = (2,1) $\rightarrow$ (1,1) and hyperfine selective collisional excitation. Its 
   main limitation is that it does not include the effects of an IR continuum at the NH$_{3}$(J,K) = (2,1) 
   $\rightarrow$ (1,1) transition frequency. As it was stressed by SW85, its influence on the anomalies is 
   difficult to predict, but they argued that if the IR radiation field had significant influence, variations 
   in their strength with the distance from the exciting source would be seen, which is not supported by 
   observations, at least in the regions S106 and W48. 

   The existence of clumps with very small dispersion velocity was questioned by Gaume et al. (\cite{gau96}), 
   based on the NH$_3$ absorption spectra in the direction of the continuum source DR21. They proposed, instead, 
   inflows and outflows of matter in different regions of the cloud. However, the mechanism producing the non-LTE 
   level population is the same as in Matsakis et al. (\cite{mat77}), that is, the selective trapping of photons 
   in the NH$_{3}$(J,K) = (2,1) $\rightarrow$ (1,1) transition. Since this model was not developed quantitatively, 
   it is not clear if it will work to explain the observed anomalies.

   Although our observations do not have high angular resolution, our high quality spectra allow the identification 
   of three distinct sources. The first one is NGC\,6334\,I, a compact source mapped in NH$_3$(J,K) = (1,1) 
   transition by JHH88, reveling the presence of a possible bipolar molecular outflow in NH$_3$ and CO 
   (Bachiller \& Cernicharo \cite{bace90}), having also a cluster of IR sources (Tapia et al. \cite{tap96}). 
   The second region is NGC\,6334\,I(N)w, which also presents a bipolar molecular outflow in SiO (Megeath \& 
   Tieftrunk \cite{meti99}), is the most intense ammonia source in the sky, but is not spatially resolved in 
   our observations. It is formed by clumps with diameters of 0.007 pc, masses of 0.2 M$_{\sun}$, kinetic temperatures 
   of 24 K and with a total mass of about 7600 M$_{\sun}$. This result suggests larger masses for this region than 
   estimated by Kuiper et al. (\cite{kui95}), assuming LTE conditions. It also exihibits velocity gradients almost 
   perpendicular to the Galactic plane, and line-width gradients parallel to the Galactic plane, showing larger 
   line-widths towards the more evolved sources. Finally, the region NGC\,6334\,I(N)e is formed by clumps with physical 
   conditions similar to those observed in I(N)w, and spread through an area larger than the telescope beam.

   The ammonia abundances in both regions are similar. Their values indicate that the regions have either ages 
   smaller than 10$^6$ years or that depletion effects are important. 

   Among the three regions identified, NGC\,6334\,I is the most evolved while NGC\,6334\,I(N)e is in a very early star 
   formation stage.

   Comparison between physical parameters determined under LTE and non-LTE conditions showed that the NH$_3$ column 
   density and ammonia abundance are the only quantities which are similar between them. Other parameters, as 
   rotational and kinetic temperatures, exhibit differences that can reach a factor of three.

\begin{acknowledgements}
      This work was supported by the Brazilian Agencies FAPESP and CNPq. CRAAE is formed by agreement 
      between INPE, UNICAMP, USP and University of Mackenzie. We would like to thank the referee, Dr. 
      E. A. Bergin, for careful readings of the manuscript and for useful suggestions.
\end{acknowledgements}


\begin{thebibliography}{}
 
   \bibitem[1992]{abko92} Abraham, Z., Kokubun, F., 1992, 
      A\&A, 257, 831. 

   \bibitem[1990]{bace90} Bachiller, R., Cernicharo, J., 1990, 
      A\&A, 239, 276.

   \bibitem[1983]{bat83} Batrla, W., Wilson, T. L., Bastien, P., Ruf, K., 1983, 
      A\&A, 128, 279.

   \bibitem[1995]{ber95} Bergin, E. A., Langer, W. D., Goldsmith, P. F., 1995, 
      ApJ, 441, 222.

   \bibitem[1978]{che78} Cheung, L., Frogel, J. A., Gezari, D. Y., Hauser, M. G., 1978, 
      ApJ, 226, L149.

   \bibitem[2000]{cumc00} Curry, C. L., Christopher, F. M., 2000, 
      ApJ, 528, 734.

   \bibitem[1988]{dan88} Danby, G., Flower, D. R., Valiron, P., Schilke, P., 
      Walmsley, C. M., 1988, MNRAS, 235, 229.

   \bibitem[1995]{dep95} De Pree, C. G., Rodríguez, L. F., Dickel, H. R., Goss, W. M., 1995, 
      ApJ, 447, 220.

   \bibitem[1987]{for87} Forster, J. R., Whiteoak, J. B., Gardner, F. F., 1987, 
      Astron. Soc. Austr., 7 (2), 189.

   \bibitem[1989]{foca89} Forster, J. R., Caswell, J. L., 1989, 
      A\&A, 213, 339.

   \bibitem[1987]{gamu87} Gaume, R. A., Mutel, R. L., 1987, 
      ApJS, 65, 193.

   \bibitem[1996]{gau96} Gaume, R. A., Wilson, T. L., Johnston, K. J., 1996, 
      ApJ, 457, L47.

   \bibitem[1982]{gez82} Gezari, D. Y., 1982, 
      ApJ, 259, L29.

   \bibitem[1993]{hahe93} Hasegawa, T. I., Herbst, E., 1993, MNRAS, 261, 83.

   \bibitem[1978]{hay78} Haynes, R. F., Caswell, J. L., Simons, L. W. J., 1978, 
      Aust. J. Phys. Astrophys. Suppl., 45, 1.

   \bibitem[1983]{hoto83} Ho, P. T. P., Townes, C. H., 1983, ARA\&A, 21, 239.

   \bibitem[1988]{jac88} Jackson, J.M., Ho, P. T. P., Haschick, A.D., 1988, 
      ApJ, 333, L73. (JHH88)

   \bibitem[1998]{kosl98} Kogan, L., Slysh, V., 1998, 
      ApJ, 497, 800.

   \bibitem[1995]{krja95} Kraemer, E. K., Jackson, J. M., 1995, 
      ApJ, 439, L9.

   \bibitem[1999]{krja99} Kraemer, K. E., Jackson, J. M., 1999, 
      ApJS, 124, 439.

   \bibitem[1995]{kui95} Kuiper, T. B. H., Peters, W. L. III, Foster, J. R., 
      Gardner, F. F., Whiteoak, J. B., 1995, ApJ, 446, 692.

   \bibitem[1986]{lou86} Loughran, L., McBreen, B., Fazio, G. G., Rengarajan, T. N., 
      Maxson, G. H., Serio, S., Sciortino, S., Ray, T. P., 1986, ApJ, 303, 629.

   \bibitem[1977]{mat77} Matsakis, D. N., Brandshaft, D., Chui, M. F., Cheung, A. C., 
      Yngvesson, K. S., Cardiasmenos, A. G., Shanley, J. F., Ho, P. T. P., 1977, 
         ApJ, 214, L67.

   \bibitem[1979]{mcb79} McBreen, B., Fazio, G. G., Stier, M., 1979, 
     ApJ, 232, L183.  

   \bibitem[1999]{meti99} Megeath, S. T., Tieftrunk, A. R., 1999, 
     ApJ, 526, L113.

   \bibitem[1989]{meba89} Menten, K. M., Batrla, W., 1989, 
     ApJ, 341, 839.

   \bibitem[1980]{moro80} Moran, J. M., Rodr\'\i guez, L. F., 1980, 
     ApJ, 236, L159.

   \bibitem[1978]{nec78} Neckel, T., 1978,
      A\&A 69, 51

   \bibitem[1996]{per96} Persi, P., Roth, M., Tapia, M., Marenzi, A. R., Felli, M., 
      Testi, L., Ferrari-Toniolo, M., 1996, A\&A, 307, 591. 

   \bibitem[1982]{rod82} Rodr\'\i guez, L. F., Cant\'o , J., Moran, J. M., 1982, 
      ApJ, 255, 103.

   \bibitem[1982]{scpu82} Scalo, J. M., Pumphrey, W. A., 1982, 
      ApJ, 258, L29.

   \bibitem[1978]{sch78} Schwartz, P. R., Bologna, J. M., Waak, J. A., 1978, 
      ApJ, 226, 469.

   \bibitem[1970]{shgo70} Shaver, P. A., Goss, W. M., 1970, 
      Aust. J. Phys. Astron. Suppl. Ser., 14, 133.

   \bibitem[1982]{stu82} Stutzki, J., Ungerechts, H.,Winnewisser, G., 1982, 
      A\&A, 111, 201. 

   \bibitem[1984]{stu84} Stutzki, J., Jackson, J. M., Olberg, M., Barrett, A. H., 
      Winnewisser, G., 1984, A\&A, 139, 258.

   \bibitem[1985]{stwi85} Stutzki, J., Winnewisser, G., 1985, 
      A\&A, 144, 13. (SW85) 

   \bibitem[1992]{suz92} Suzuki, H., Yamamoto, S., Ohishi, M., Kaifu, M., Ishikawa, S., 
      Hirhara, Y., Takano, S., 1992, ApJ, 392, 551. 

   \bibitem[1996]{tap96} Tapia, M., Persi, P., Roth, M., 1996, 
      A\&A, 316, 102.

   \bibitem[2000]{viab00} Vilas-Boas, J. W. S., Abraham, Z., 2000, 
      A\&A, 355, 1115.

   \bibitem[1988]{vil88} Vilas-Boas, J. W. S., Scalise, E.J., Monteiro do Vale, J.L.,
      1988, Ap\&SS, 141, 339.

   \bibitem[1983]{waun83} Walmsley, C. M. e Ungerechts, H., 1983, A\&A, 122, 164.

   \bibitem[1998]{wal98} Walsh, A. J., Burton, M. G., Hyland, A. R., Robinson, G.,
      1998, MNRAS, 301, 640.

\end{thebibliography}
\end{document}